\documentclass[aps,prd,twocolumn,showpacs,superscriptaddress,nofootinbib]{revtex4-1}
\pdfoutput=1

\usepackage{graphicx}
\usepackage{bm}
\usepackage{amssymb,amsmath,latexsym}
\usepackage{color}
\definecolor{darkblue}{RGB}{0,0,196}
\definecolor{darkred}{RGB}{196,0,0}
\usepackage[colorlinks=true,linktocpage=true,linkcolor=darkblue,citecolor=darkblue,urlcolor=darkblue]{hyperref}
\usepackage{array}

\def\be{\begin{equation}}
\def\ee{\end{equation}}
\def\ba{\begin{eqnarray}}
\def\ea{\end{eqnarray}}

\def\be{\begin{eqnarray}}
\def\ee{\end{eqnarray}}

\def\[{\left[}
\def\]{\right]}  

\newcolumntype{P}[1]{>{\centering\arraybackslash}p{#1}}
\begin{document}

\title{Next-to-next-to leading-order hard-thermal-loop perturbation-theory predictions for the curvature of the QCD phase transition line}

\author{Najmul Haque} 
\affiliation{School of Physical Sciences, National Institute of Science Education and Research, HBNI, Jatni 752050, India}

\author{Michael Strickland} 
\affiliation{Department of Physics, Kent State University, Kent, OH 44242, USA}

\begin{abstract}
We present predictions for the second- and fourth-order curvature coefficients of the QCD phase transition line using the NNLO HTLpt-resummed thermodynamic potential.  We present three cases corresponding to (i) $\mu_s = \mu_l = \mu_B/3$, (ii) $\mu_s=0$, $\mu_l = \mu_B/3,$ and (iii) $S = 0$, $Q/B = 0.4$, $\mu_l = \mu_B/3$.  In all three cases, we find excellent agreement with continuum extrapolated lattice QCD results for $\kappa_2$, given current statistical uncertainties. We also make HTLpt predictions for $\kappa_4$ in all three cases, finding again excellent agreement with lattice extractions of this coefficient where available.
\end{abstract}

\date{\today}



\maketitle


It is well established that, at low energy density, quarks and gluons are confined within hadrons via the long-range strong interactions present in Quantum Chromodynamics (QCD) and that there is a large chiral condensate.  At high temperatures and low net baryon density, numerical lattice QCD calculations find that nuclear matter becomes deconfined and chiral symmetry is restored.  The resulting phase diagram of QCD encodes the temperature and chemical-potential dependence of these transitions, including the order of each phase transition. The nature of the QCD phase diagram at finite temperature and baryon density, including the dependence of the phase transition temperature on the net baryon density, has been the subject of very active research in recent years~\cite{Cea:2015cya,Bonati:2015bha,Bonati:2018nut,Bonati:2014rfa,Borsanyi:2020fev,Bazavov:2018mes,Toublan:2004ks,Endrodi:2011gv,Bellwied:2015rza,Giordano:2020uvk,Giordano:2020roi,Scherzer:2020kiu,Attanasio:2020spv,Pasztor:2020dur,Fischer:2012vc,Fischer:2014ata,Isserstedt:2019pgx,Bratovic:2012qs,Kovacs:2016juc,Fu:2019hdw,Gao:2020fbl,Gao:2020qsj,Andronic:2005yp,Becattini:2012xb,Alba:2014eba,Vovchenko:2015idt,Adamczyk:2017iwn,Bluhm:2020rha}. 

Two regions of the QCD phase diagram are of particular interest, namely, (i) large net baryon density and vanishing temperature and (ii) large temperature and vanishing net baryon density.  At large to moderate values of the net baryon density, it is expected that confined nuclear matter becomes deconfined via a first order phase transition. If one moves along the phase transition line from high to low values of the net baryon density, this first order phase transition line ends at the QCD critical point and, beyond this point, there is a smooth crossover at small-to-vanishing net baryon density. The large temperature and small net baryon density region of the QCD phase diagram can be accessed via various heavy-ion experiments. Currently, the Large Hadron Collider at CERN~\cite{Aamodt:2010pa,Aamodt:2010jd,Chatrchyan:2011sx,Aamodt:2010cz,Aamodt:2010pb,ALICE:2011ab,Aamodt:2008zz} is focused on the study of deconfined matter at small net baryon densities, whereas the Beam-Energy-Scan program at the Relativistic Heavy Ion Collider~\cite{Mohanty:2011nm} in New York is currently focusing on the nature of the transition at larger net baryon density. This latter program will be complemented by future facilities at the Facility for Antiproton and Ion Research~\cite{Wilczek:2010ae} in Darmstadt, Germany and the Joint Institute Nuclear Research~\cite{Sorin:2011zz} in Dubna, Russia.  

In parallel with the worldwide experimental program, theorists have used various techniques to determine the curvature of the QCD phase transition line in the $T$-$\mu_B$ plane, where $\mu_B$ is the baryochemical potential associated with a given net baryon density. Various methods have been used to extract the curvature of the QCD transition line, including but not limited to direct numerical lattice QCD calculations~\cite{Cea:2015cya,Bonati:2015bha,Bonati:2018nut,Bonati:2014rfa,Borsanyi:2020fev,Bazavov:2018mes,Toublan:2004ks,Endrodi:2011gv,Bellwied:2015rza}, Dyson-Schwinger-Equation approaches~\cite{Fischer:2012vc,Fischer:2014ata,Isserstedt:2019pgx}, the Polyakov-loop improved Nambu-Jona-Lasino model~\cite{Bratovic:2012qs,Kovacs:2016juc}, functional renormalization Group approaches~\cite{Fu:2019hdw,Gao:2020fbl,Gao:2020qsj}, and phenomenological extractions of the freeze-out temperature using thermal models~\cite{Andronic:2005yp,Becattini:2012xb,Alba:2014eba,Vovchenko:2015idt,Adamczyk:2017iwn,Bluhm:2020rha}. In this brief report we calculate the curvature of the QCD phase transition line using next-to-next-to leading-order (NNLO or three-loop) hard-thermal-loop perturbation theory (HTLpt).  

For the purposes of this work, we make use of the temperature and chemical-potential dependence of the NNLO HTLpt pressure which was calculated analytically in Ref.~\cite{Haque:2014rua}.  This NNLO HTLpt calculation showed good agreement between the resummed perturbative calculations and continuum extrapolated lattice results for a wide array of lattice observables, including the pressure versus temperature and various susceptibilities.  The success of HTLpt can be attributed to the fact that shifting the starting point for the finite-temperature and/or density loop expansion builds fundamental classical physics related to plasma screening and damping into the calculation, thereby resumming a large class of diagrams that are important in a high temperature and/or density QGP and curing all infrared problems in the chromoelectric sector~\cite{Andersen:1999fw,Andersen:1999va,Andersen:2000zn,Andersen:2000yj,Andersen:2002ey,Andersen:2001ez,Andersen:2003zk,Andersen:2004fp,Andersen:2009tw,Andersen:2009tc,Andersen:2010ct,Andersen:2010wu,Andersen:2011sf,Andersen:2011ug,Haque:2012my,Mogliacci:2013mca,Haque:2013qta,Haque:2013sja,Andersen:2015eoa,Du:2020odw}.

Herein, we compare NNLO HTLpt calculations of the curvature of the QCD phase transition line with available lattice data.  Typically, lattice studies are restricted to the region of small \mbox{$\mu_B$}.  In the case of the HTLpt predictions, however, we find that the typical quartic fit form used to extract the curvature coefficients works well for $\mu_B/T \lesssim 1$.  For our main results, we will compare the curvature coefficients extracted in three cases that have been considered in the lattice literature, namely (i) $\mu_s = \mu_l = \mu_B/3$, (ii) $\mu_s = 0, \mu_l = \mu_B/3$, and (iii) $S = 0, Q/B=0.4, \mu_l = \mu_B/3$, where $\mu_s$ and $\mu_l$ are the strange and light quark chemical potentials, respectively.  In the third case, which most closely mimics real-world conditions generated in Au-Au as well as Pb-Pb collisions (as the atomic number to mass number ratio is $\sim 0.4$ for these nuclei), the strange quark chemical potential is a function of $T$ and $\mu_B$, which guarantees that the net strangeness ($S$) is zero and that the charge to baryon number ratio ($Q/B$) is held fixed.  We find that the existing analytic NNLO HTLpt result of Ref.~\cite{Haque:2014rua} results in very good agreement between resummed perturbative QCD and the world's lattice data for the curvature coefficients $\kappa_2$ and $\kappa_4$ in all three cases.  
 
Our brief report is structured as follows.  We first present a brief overview of the HTLpt formalism. We then present our NNLO HTLpt results for the second- and fourth-order curvatures and compare to existing lattice results for $\kappa_2$ and $\kappa_4$. Finally, we summarize our findings.  

{\bf HTLpt formalism} -- The Minkowski-space QCD Lagrangian density with massless quarks is
\ba
{\cal L}_{\rm QCD}&=&
-\frac{1}{2}{\rm Tr}\left[G_{\mu\nu}G^{\mu\nu}\right]
+\sum_i\bar{\psi}_i \left[i\gamma^\mu D_\mu -\gamma_0\mu_i\right] \psi_i
\nonumber\\
&&+\ {\cal L}_{\rm gf}
+{\cal L}_{\rm ghost} + \Delta{\cal L}_{\rm QCD}\;,
\label{qcd_lag}
\ea
where $\Delta{\cal L}_{\rm QCD}$ contains the  necessary counterterms that cancel the vacuum ultraviolet divergences in perturbative calculations. The gluon field strength is $G_{\mu\nu} = \partial_{\mu}A_{\nu}-\partial_{\nu}A_{\mu} -ig[A_{\mu},A_{\nu}]$ and $D_\mu = \partial_\mu - i g A_\mu$ is the
covariant derivative in the fundamental representation. Additionally, in the quark sector $\mu_i$ represents the chemical potential of the $i^{\mbox{th}}$ flavor and there is an explicit sum over the $N_f$ quark flavors, whereas ${\cal L}_{\rm ghost}$ represents the ghost term that depends on the choice of the gauge-fixing term ${\cal L}_{\rm gf}$.

HTL perturbation theory (HTLpt) is a reorganization of bare perturbation theory for QCD at finite temperature and chemical potential(s). The HTLpt Lagrangian density is obtained as  \cite{Andersen:2003zk}
\begin{eqnarray}
\mathcal{L}=\left.\left(\mathcal{L}_{\mathrm{QCD}}+\mathcal{L}_{\mathrm{HTL}}\right)\right|_{g \rightarrow \sqrt{\delta} g}+\Delta \mathcal{L}_{\mathrm{HTL}} \, ,
\end{eqnarray}
where $\mathcal{L}_{\mathrm{QCD}}$ is the QCD Lagrangian density~\eqref{qcd_lag} and $\mathcal{L}_{\mathrm{HTL}} $ represents the HTL improvement term which can be written as~\cite{Braaten:1991gm,Andersen:2002ey,Andersen:2003zk}
\begin{eqnarray}
\mathcal{L}_{\mathrm{HTL}} &=& (1-\delta) i m_{q}^{2} \bar{\psi} \gamma^{\mu}\left\langle\frac{y_{\mu}}{y \cdot D}\right\rangle_{\hat{\mathbf{y}}} \psi
\nonumber \\ 
&&-\frac{1}{2}(1-\delta) m_{D}^{2} \operatorname{Tr}\left[G_{\mu \alpha}\left\langle\frac{y^{\alpha} y_{\beta}}{(y \cdot D)^{2}}\right\rangle_{\hat{\mathbf{y}}} G^{\mu \beta}\right] \! , \;\;\;
\label{htl_lag}
\end{eqnarray}
where the first term is due to the quark sector and the second term is due to the  gluon and ghost sector. In the quark sector, $D_\mu$ is the covariant derivative in fundamental representation whereas, in the gluon sector,  $D_\mu$ represents the covariant derivative in the adjoint representation. The four-vector $y^\mu\equiv (1,\hat{\textbf{y}})$ is a light-like four-vector which encodes the velocity of the hard quarks and gluons. The angular bracket in Eq.~\eqref{htl_lag} indicates an average over the direction of $\hat{\textbf{y}}$ and the parameter $\delta$ is the formal expansion parameter in HTLpt. To calculate the thermodynamic potential, we should truncate the relevant expression at some specific order in $\delta$, which depends upon the loop-order we are interested in and in the end, one should set $\delta=1$.  Note that if one sets $\delta=1$ in Eq.~\eqref{htl_lag}, one gets back QCD Lagrangian. 

The quantities $m_q$ and $m_D$ can be identified with the thermal quark mass and Debye screening mass, respectively. In principle, these are unknown parameters and can be determined in HTLpt by a variational prescription or using effective field theory methods. As in the case of vacuum QCD, the HTLpt expansion produces ultraviolet divergences. In bare perturbation theory,  the ultraviolet divergences can be renormalized by the counterterm $\Delta\mathcal{L}_\text{QCD}$. Although there is not yet a general proof, it has been shown through three-loop order that all ultraviolet divergences not removed by $\Delta\mathcal{L}_\text{QCD}$ can be removed using simple mass and coupling constant renormalizations, which can be collected in $\Delta\mathcal{L}_\text{HTL}$~\cite{Haque:2014rua}.  

{\bf Extracting the curvature of the QCD transition line} --  At small baryochemical potential, the chemical potential dependence of the crossover line can be expressed as
\begin{eqnarray}
\frac{T_{c}^\mu}{T_{c}^0}&=&1-\kappa_{2}\left(\frac{\mu_{B}}{T_{c}^\mu}\right)^{2}-\kappa_{4}\left(\frac{\mu_{B}}{T_{c}^\mu}\right)^{4}\nonumber\\ &&\hspace{2cm}-\kappa_{6}\left(\frac{\mu_{B}}{T_{c}^\mu}\right)^{6} -\;  \ldots,
\label{eq:tcmuexp}
\end{eqnarray}
where $T_c^\mu$ is the chemical potential dependent crossover temperature and $T_c^0$ is the crossover temperature at \mbox{$\mu_B=0$}.
\begin{figure}[b]
	\centering
	\includegraphics[width=\linewidth]{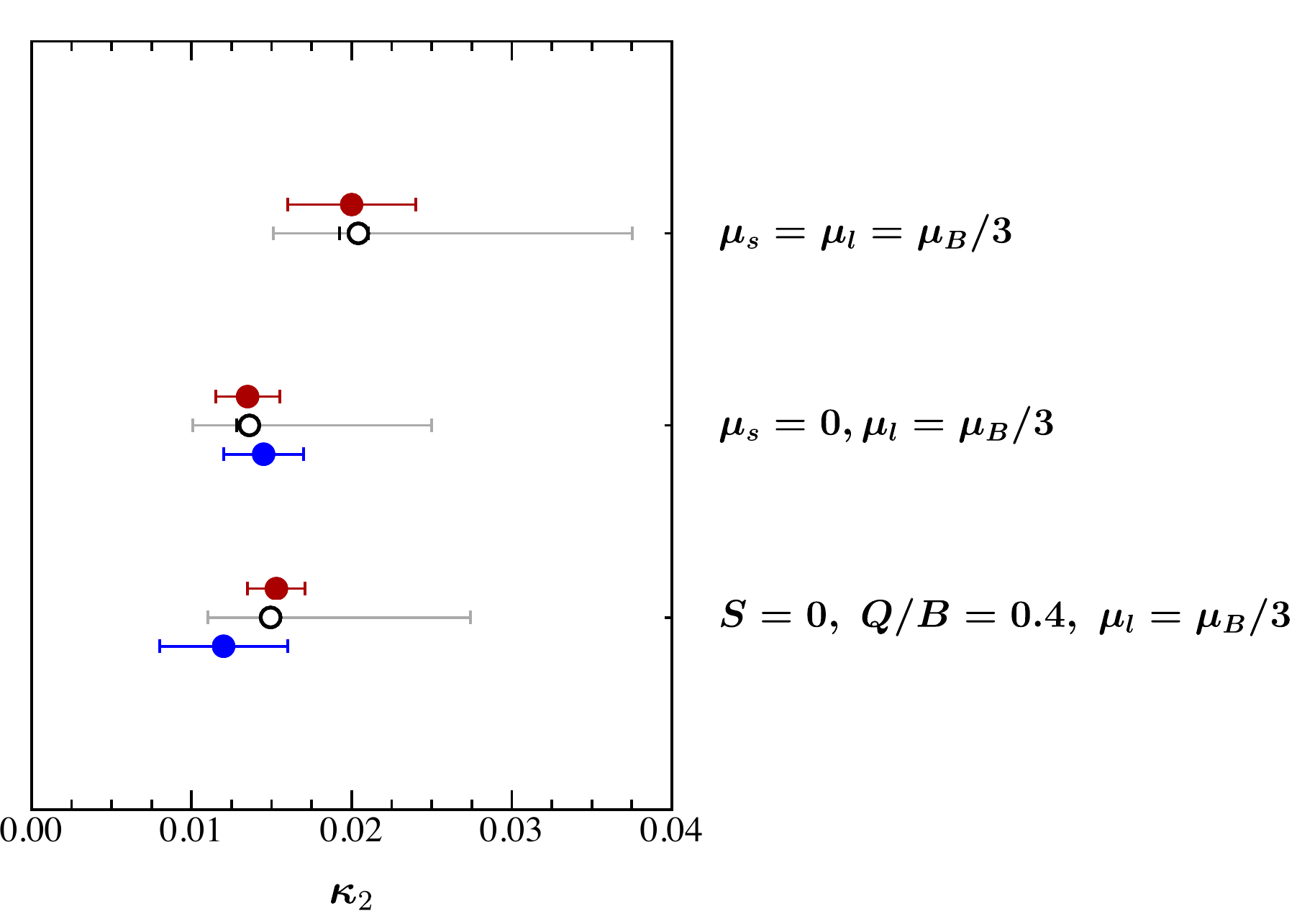}
	\caption{Filled circles are lattice calculations of $\kappa_2$~\cite{Cea:2015cya,Bonati:2015bha,Bonati:2018nut,Borsanyi:2020fev,Bazavov:2018mes}, from top to bottom, respectively.  Red filled circles are results obtained using the imaginary chemical potential method and blue filled circles are results obtained using Taylor expansions around $\mu_B=0$.  Black open circles are the NNLO HTLpt predictions. The black error bars associated with the HTLpt predictions result from variation of the assumed renormalization scale and the gray error bars are the result of varying the thermodynamic variable used to determine the curvature.}
	\label{fig:kappa2}
\end{figure}

\begin{figure}[t]
	\centering
	\includegraphics[width=\linewidth]{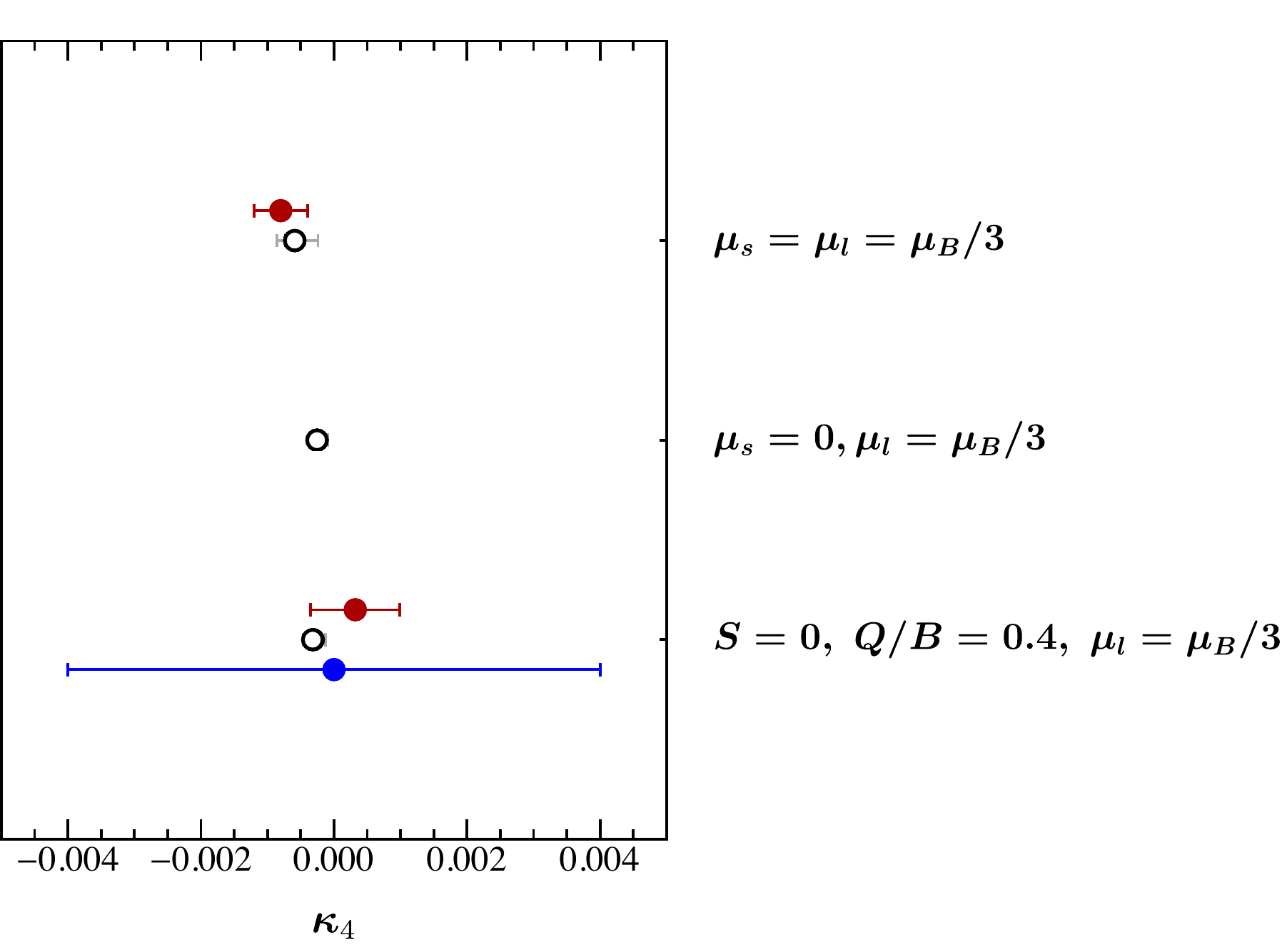}
	\caption{Filled circles are lattice calculations of $\kappa_4$ from Refs.~\cite{Borsanyi:2020fev,Bazavov:2018mes,Bonati:2014rfa}, from top to bottom, respectively.  The color coding etc. for the symbols and error bars is the same as in Fig.~\ref{fig:kappa2}.}
	\label{fig:kappa4}
\end{figure}

In order to extract a phase transition temperature in HTLpt one can extract the temperature at which the NNLO HTLpt resummed pressure goes to zero. Below this temperature, $T_\text{P=0}^\mu$, the QGP phase is unstable. In practice, this provides a lower limit on the phase transition temperature since the crossover to a hadron gas occurs before the deconfined QGP phase pressure goes to zero~\footnote{Note that in lattice calculations one uses the peak in the chiral susceptibility to determine the pseudocritical temperature.  This has no obvious connection to the point $P=0$ in HTLpt.}. To obtain the HTLpt estimate for the pressure, we use the NNLO HTLpt expression for the pressure contained in Eq.~(4.7) of Ref.~\cite{Haque:2014rua}.  For the Debye mass, thermal quark masses, and one-loop running coupling constant, we use the same prescriptions as Ref.~\cite{Haque:2014rua}. The final NNLO HTLpt result depends on the temperature $T$, the individual quark chemical potentials $\mu_i$, and the renormalization scales for the quark and gluon sectors, $\Lambda_q$ and $\Lambda_g$.  The central values for these two scales are taken to be $\Lambda_g^0 = 2 \pi T$ and $\Lambda_q^0 =2\pi\sqrt{T^2 + \mu^2/\pi^2}$.  To estimate the uncertainty associated with the renormalization scale choice, we vary the renormalization scales by a factor of two around the central values, i.e. $ \Lambda_g = c \Lambda_g^0$ and $ \Lambda_q = c \Lambda_q^0$, with $c \in [1/2,2]$.  The resulting NNLO HTLpt pressure can be seen in Fig.~1 of Ref.~\cite{Haque:2014rua}.  For the central values of the two scales and $\mu_i=0$, one finds that the pressure vanishes at $T_\text{P=0}^\mu = 148.4$ MeV.  The bands shown in Fig.~1 of Ref.~\cite{Haque:2014rua} indicate the variation of the pressure under $c \in [1/2,2]$.

The choice of $T_\text{P=0}^\mu$ is not unique and one could also consider using the curvature of lines of constant energy or entropy density.  To estimate this uncertainty, in our final results we use the values of the energy and entropy density corresponding to the point at which $P=0$ for $\mu_i=0$.  We then determine lines of constant energy and entropy density using these values. In all three cases, we compute the curvatures in two manners:  for each choice of $c$ we (a) numerically solve for $T_\text{P=0}^\mu$, $T_\text{E=const}^\mu$, and $T_\text{s=const}^\mu$ and then fit $\kappa_2$ and $\kappa_4$ using a polynomial fit (using 100 points in the range $0 \leq \mu_B \leq 100$ MeV) to the form given in Eq.~\eqref{eq:tcmuexp} and (b) we analytically take derivatives with respect to $\mu_B$ of the NNLO thermodynamic variables and use the resulting expressions.  We find that the values coincide within numerical uncertainties, except in case (iii) in which uncertainties due to the $\mu_s$ interpolating function become important resulting in a difference of approximately 30\%.  Since the derivative method is generally more reliable, we report the results obtained using this method in the tables and figures in case (iii).

Since the point $P=0$ corresponds to the place where the high-temperature phase becomes unstable, this measure is the most physical motivated definition of the phase transition temperature in HTLpt.  We call this prescription ``HTLpt I'' in Tables \ref{tab:k2comp} and \ref{tab:k4comp} and use this value as the central value in all of our figures and tables.  For the ``HTLpt I'' case, we estimate the uncertainty in this central value by varying $c \in [1/2,2]$ and indicate this with a black error bar in the figures. To quantify the uncertainty coming from the choice of constant $P$, $E$, and $s$ curvatures, we present the variation in this case as ``HTLpt II'' in Tables \ref{tab:k2comp} and \ref{tab:k4comp} and visually indicate this variation with gray error bars in the figures. 

We perform our analysis in three distinct physical cases corresponding to
\ba
\text{(i)} &~~~& \mu_s = \mu_l = \mu_B/3 \, , \nonumber \\
\text{(ii)} &~~~& \mu_s = 0, \mu_l = \mu_B/3 \, , \nonumber \\
\text{(iii)} &~~~& S = 0, Q/B=0.4, \mu_l = \mu_B/3 \, .
\ea

\begin{widetext}
	
\setlength{\tabcolsep}{-1pt}
\begin{table}[b]
\begin{center}
		\caption{Comparison of NNLO HTLpt predictions for $\kappa_2$ with lattice results.  See text for description of the HTLpt I and II columns.}
\begin{tabular}{l l l l }
\hline
\hline
~ {\bf Case} ~  & {\bf ~Lattice $\kappa_2$~} \hspace*{4cm}& {\bf ~HTLpt $\kappa_2$~ I}  \hspace{3cm} &{\bf ~HTLpt $\kappa_2$~ II}  \hspace{3cm}\\
\hline
~ $\mu_s=\mu_l=\mu_B/3$ ~ & ~ 0.020(4)~\cite{Cea:2015cya} ~ & ~ $0.0204_{-0.0006}^{+0.0012}$ ~ & $0.0204^{+0.0171}_{-0.0053}$ \\
~  $\mu_s$ = 0, $\mu_l=\mu_B/3$ ~  & ~ 0.0135(20)~\cite{Bonati:2015bha} ,  0.0145(25)~\cite{Bonati:2018nut} ~ & ~ $0.0136_{-0.0004}^{+0.0008}$ ~ &  $0.0136_{-0.0036}^{+0.0114}$ \\
~ $S=0$, $Q/B=0.4$,\ $\mu_l=\mu_B/3$ ~ & ~ 0.0153(18)~\cite{Borsanyi:2020fev} ,  0.012(4)~\cite{Bazavov:2018mes} & ~$0.0149^{+0.0005}_{-0.0002}$ &  $0.0149^{+0.0125}_{-0.0039}$\\
 \hline
  \hline
\end{tabular}
\end{center}

\label{tab:k2comp}
\end{table}


\setlength{\tabcolsep}{7pt}
\begin{table}[t]
\begin{center}
\caption{Comparison of NNLO HTLpt predictions for $\kappa_4$ with lattice results.}
\begin{tabular}{l l l l}
\hline
\hline
~ {\bf Case} ~  & {\bf ~Lattice $\kappa_4$~} & {\bf ~HTLpt $\kappa_4$~ I}& {\bf ~HTLpt $\kappa_4$~ II} \\
\hline
 ~$\mu_s=\mu_l =\mu_B/3$  & $-0.0008(4)$~\cite{Bonati:2014rfa} &  $-0.000594_{-0.000139}^{+0.000015}$ ~ &  $-0.000594_{-0.000267}^{+0.000349}$ ~ \\
~  $\mu_s$ = 0, $\mu_l=\mu_B/3$ ~  & ~ -- ~ &  $-0.000255_{-0.000056}^{+0.000004}$ ~ &$-0.000255_{-0.000120}^{+0.000157}$\\
~ $S=0$, $Q/B=0.4$,  $\mu_l=\mu_B/3$ ~ &  0.00032(67)~\cite{Borsanyi:2020fev} , 0.000(4)~\cite{Bazavov:2018mes} & $-0.000312$ & $-0.000312^{+0.000188}_{-0.000144}$ \\
 \hline
 \hline
\end{tabular}

\end{center}

\label{tab:k4comp}
\end{table}
\end{widetext}

The last case corresponds to the physical case with fixed charge to baryon number and zero net strangeness.  In this case, $\mu_s$ is a nontrivial function of $\mu_B$ and $T$.  Herein, we use lattice data provided by the authors of Ref.~\cite{Borsanyi:2020fev} for case (iii).  Our results in these three cases are presented in Figs.~\ref{fig:kappa2} and \ref{fig:kappa4}. Our results are compared with lattice QCD results obtained using different indirect methods because the finite chemical potential is not directly accessible numerically on the lattice due to the sign problem.  Refs.~\cite{Bonati:2015bha,Cea:2015cya,Borsanyi:2020fev} used the imaginary chemical potential method and Refs.~\cite{Bonati:2018nut,Bazavov:2018mes} used the Taylor expansion method to measure $\kappa_2$.  Refs.~\cite{Borsanyi:2020fev,Bazavov:2018mes,Bonati:2014rfa} also reported measurements for $\kappa_4$. In Ref.~\cite{Bonati:2014rfa}, $\kappa_4$ is calculated at imaginary chemical potential and the corresponding value in table~\ref{tab:k4comp} is obtained from the renormalized chiral condensate. These are all collected in Figs.~\ref{fig:kappa2} and~\ref{fig:kappa4} and Tables~\ref{tab:k2comp} and~\ref{tab:k4comp}. As these Figures and Tables demonstrate, we find excellent agreement between the NNLO HTLpt results and lattice data.  

Note that the lattice results for $\kappa_4$ are consistent with zero within uncertainties with differing signs for the central values; whereas in HTLpt, we find it to be negative but small in all cases. This provides some motivation for improving the accuracy of the lattice analyses of $\kappa_4$ in order to check whether HTLpt is reliable in this case.  Finally, we note that we also extracted $\kappa_6$ using the derivative method, finding $\kappa_6 \simeq  -2.65 \times 10^{-5}$ and $\kappa_6 \simeq  -5.2 \times 10^{-5}$ for cases (i) and (ii), respectively, when using $c=1$ in the renormalization scale.  In case (iii) the numerical uncertainties associated with the interpolating function used for $\mu_s$ become important, so we do not report a value for $\kappa_6$ for this case.

{\bf Summary} -- In this brief report, we presented NNLO HTLpt predictions for the second- and fourth-order curvatures of the QCD phase transition line and compared our results with available lattice QCD measurements of these coefficients.  Figs.~\ref{fig:kappa2} and \ref{fig:kappa4}, together with Tables \ref{tab:k2comp} and \ref{tab:k4comp}, demonstrate that NNLO HTLpt is consistent with existing lattice calculations of $\kappa_2$ and $\kappa_4$ in all three cases considered in the lattice QCD literature.  In the case of $\kappa_4$, lattice measurements only exist in two of the three cases considered.  We made predictions for the other two cases, which can be checked with future lattice calculations. Finally, we also made predictions for the sixth order curvature coefficient $\kappa_6$.

{\bf Acknowledgements} --
We thank P.~Parotto for discussions and providing the extraction of $\mu_s$ necessary for case (3).
N.H. was supported by the Department of Atomic Energy, Govt. of India and also in part by the SERB-SRG under Grant No. SRG/2019/001680. M.S. was supported by the U.S. Department of Energy, Office of Science, Office of Nuclear Physics Award No.~DE-SC0013470. 

\bibliography{curvature}

\begin{thebibliography}{60}%
\makeatletter
\providecommand \@ifxundefined [1]{%
 \@ifx{#1\undefined}
}%
\providecommand \@ifnum [1]{%
 \ifnum #1\expandafter \@firstoftwo
 \else \expandafter \@secondoftwo
 \fi
}%
\providecommand \@ifx [1]{%
 \ifx #1\expandafter \@firstoftwo
 \else \expandafter \@secondoftwo
 \fi
}%
\providecommand \natexlab [1]{#1}%
\providecommand \enquote  [1]{``#1''}%
\providecommand \bibnamefont  [1]{#1}%
\providecommand \bibfnamefont [1]{#1}%
\providecommand \citenamefont [1]{#1}%
\providecommand \href@noop [0]{\@secondoftwo}%
\providecommand \href [0]{\begingroup \@sanitize@url \@href}%
\providecommand \@href[1]{\@@startlink{#1}\@@href}%
\providecommand \@@href[1]{\endgroup#1\@@endlink}%
\providecommand \@sanitize@url [0]{\catcode `\\12\catcode `\$12\catcode
  `\&12\catcode `\#12\catcode `\^12\catcode `\_12\catcode `\%12\relax}%
\providecommand \@@startlink[1]{}%
\providecommand \@@endlink[0]{}%
\providecommand \url  [0]{\begingroup\@sanitize@url \@url }%
\providecommand \@url [1]{\endgroup\@href {#1}{\urlprefix }}%
\providecommand \urlprefix  [0]{URL }%
\providecommand \Eprint [0]{\href }%
\providecommand \doibase [0]{http://dx.doi.org/}%
\providecommand \selectlanguage [0]{\@gobble}%
\providecommand \bibinfo  [0]{\@secondoftwo}%
\providecommand \bibfield  [0]{\@secondoftwo}%
\providecommand \translation [1]{[#1]}%
\providecommand \BibitemOpen [0]{}%
\providecommand \bibitemStop [0]{}%
\providecommand \bibitemNoStop [0]{.\EOS\space}%
\providecommand \EOS [0]{\spacefactor3000\relax}%
\providecommand \BibitemShut  [1]{\csname bibitem#1\endcsname}%
\let\auto@bib@innerbib\@empty
\bibitem [{\citenamefont {Cea}\ \emph {et~al.}(2016)\citenamefont {Cea},
  \citenamefont {Cosmai},\ and\ \citenamefont {Papa}}]{Cea:2015cya}%
  \BibitemOpen
  \bibfield  {author} {\bibinfo {author} {\bibfnamefont {P.}~\bibnamefont
  {Cea}}, \bibinfo {author} {\bibfnamefont {L.}~\bibnamefont {Cosmai}}, \ and\
  \bibinfo {author} {\bibfnamefont {A.}~\bibnamefont {Papa}},\ }\href {\doibase
  10.1103/PhysRevD.93.014507} {\bibfield  {journal} {\bibinfo  {journal} {Phys.
  Rev.}\ }\textbf {\bibinfo {volume} {D93}},\ \bibinfo {pages} {014507}
  (\bibinfo {year} {2016})},\ \Eprint {http://arxiv.org/abs/1508.07599}
  {arXiv:1508.07599 [hep-lat]} \BibitemShut {NoStop}%
\bibitem [{\citenamefont {Bonati}\ \emph {et~al.}(2015)\citenamefont {Bonati},
  \citenamefont {D'Elia}, \citenamefont {Mariti}, \citenamefont {Mesiti},
  \citenamefont {Negro},\ and\ \citenamefont {Sanfilippo}}]{Bonati:2015bha}%
  \BibitemOpen
  \bibfield  {author} {\bibinfo {author} {\bibfnamefont {C.}~\bibnamefont
  {Bonati}}, \bibinfo {author} {\bibfnamefont {M.}~\bibnamefont {D'Elia}},
  \bibinfo {author} {\bibfnamefont {M.}~\bibnamefont {Mariti}}, \bibinfo
  {author} {\bibfnamefont {M.}~\bibnamefont {Mesiti}}, \bibinfo {author}
  {\bibfnamefont {F.}~\bibnamefont {Negro}}, \ and\ \bibinfo {author}
  {\bibfnamefont {F.}~\bibnamefont {Sanfilippo}},\ }\href {\doibase
  10.1103/PhysRevD.92.054503} {\bibfield  {journal} {\bibinfo  {journal} {Phys.
  Rev.}\ }\textbf {\bibinfo {volume} {D92}},\ \bibinfo {pages} {054503}
  (\bibinfo {year} {2015})},\ \Eprint {http://arxiv.org/abs/1507.03571}
  {arXiv:1507.03571 [hep-lat]} \BibitemShut {NoStop}%
\bibitem [{\citenamefont {Bonati}\ \emph {et~al.}(2018)\citenamefont {Bonati},
  \citenamefont {D'Elia}, \citenamefont {Negro}, \citenamefont {Sanfilippo},\
  and\ \citenamefont {Zambello}}]{Bonati:2018nut}%
  \BibitemOpen
  \bibfield  {author} {\bibinfo {author} {\bibfnamefont {C.}~\bibnamefont
  {Bonati}}, \bibinfo {author} {\bibfnamefont {M.}~\bibnamefont {D'Elia}},
  \bibinfo {author} {\bibfnamefont {F.}~\bibnamefont {Negro}}, \bibinfo
  {author} {\bibfnamefont {F.}~\bibnamefont {Sanfilippo}}, \ and\ \bibinfo
  {author} {\bibfnamefont {K.}~\bibnamefont {Zambello}},\ }\href {\doibase
  10.1103/PhysRevD.98.054510} {\bibfield  {journal} {\bibinfo  {journal} {Phys.
  Rev.}\ }\textbf {\bibinfo {volume} {D98}},\ \bibinfo {pages} {054510}
  (\bibinfo {year} {2018})},\ \Eprint {http://arxiv.org/abs/1805.02960}
  {arXiv:1805.02960 [hep-lat]} \BibitemShut {NoStop}%
\bibitem [{\citenamefont {Bonati}\ \emph {et~al.}(2014)\citenamefont {Bonati},
  \citenamefont {D'Elia}, \citenamefont {Mariti}, \citenamefont {Mesiti},
  \citenamefont {Negro},\ and\ \citenamefont {Sanfilippo}}]{Bonati:2014rfa}%
  \BibitemOpen
  \bibfield  {author} {\bibinfo {author} {\bibfnamefont {C.}~\bibnamefont
  {Bonati}}, \bibinfo {author} {\bibfnamefont {M.}~\bibnamefont {D'Elia}},
  \bibinfo {author} {\bibfnamefont {M.}~\bibnamefont {Mariti}}, \bibinfo
  {author} {\bibfnamefont {M.}~\bibnamefont {Mesiti}}, \bibinfo {author}
  {\bibfnamefont {F.}~\bibnamefont {Negro}}, \ and\ \bibinfo {author}
  {\bibfnamefont {F.}~\bibnamefont {Sanfilippo}},\ }\href {\doibase
  10.1103/PhysRevD.90.114025} {\bibfield  {journal} {\bibinfo  {journal} {Phys.
  Rev.}\ }\textbf {\bibinfo {volume} {D90}},\ \bibinfo {pages} {114025}
  (\bibinfo {year} {2014})},\ \Eprint {http://arxiv.org/abs/1410.5758}
  {arXiv:1410.5758 [hep-lat]} \BibitemShut {NoStop}%
\bibitem [{\citenamefont {Borsanyi}\ \emph {et~al.}(2020)\citenamefont
  {Borsanyi}, \citenamefont {Fodor}, \citenamefont {Guenther}, \citenamefont
  {Kara}, \citenamefont {Katz}, \citenamefont {Parotto}, \citenamefont
  {Pasztor}, \citenamefont {Ratti},\ and\ \citenamefont
  {Szabo}}]{Borsanyi:2020fev}%
  \BibitemOpen
  \bibfield  {author} {\bibinfo {author} {\bibfnamefont {S.}~\bibnamefont
  {Borsanyi}}, \bibinfo {author} {\bibfnamefont {Z.}~\bibnamefont {Fodor}},
  \bibinfo {author} {\bibfnamefont {J.~N.}\ \bibnamefont {Guenther}}, \bibinfo
  {author} {\bibfnamefont {R.}~\bibnamefont {Kara}}, \bibinfo {author}
  {\bibfnamefont {S.~D.}\ \bibnamefont {Katz}}, \bibinfo {author}
  {\bibfnamefont {P.}~\bibnamefont {Parotto}}, \bibinfo {author} {\bibfnamefont
  {A.}~\bibnamefont {Pasztor}}, \bibinfo {author} {\bibfnamefont
  {C.}~\bibnamefont {Ratti}}, \ and\ \bibinfo {author} {\bibfnamefont {K.~K.}\
  \bibnamefont {Szabo}},\ }\href {\doibase 10.1103/PhysRevLett.125.052001}
  {\bibfield  {journal} {\bibinfo  {journal} {Phys. Rev. Lett.}\ }\textbf
  {\bibinfo {volume} {125}},\ \bibinfo {pages} {052001} (\bibinfo {year}
  {2020})},\ \Eprint {http://arxiv.org/abs/2002.02821} {arXiv:2002.02821
  [hep-lat]} \BibitemShut {NoStop}%
\bibitem [{\citenamefont {Bazavov}\ \emph {et~al.}(2019)\citenamefont {Bazavov}
  \emph {et~al.}}]{Bazavov:2018mes}%
  \BibitemOpen
  \bibfield  {author} {\bibinfo {author} {\bibfnamefont {A.}~\bibnamefont
  {Bazavov}} \emph {et~al.} (\bibinfo {collaboration} {HotQCD}),\ }\href
  {\doibase 10.1016/j.physletb.2019.05.013} {\bibfield  {journal} {\bibinfo
  {journal} {Phys. Lett. B}\ }\textbf {\bibinfo {volume} {795}},\ \bibinfo
  {pages} {15} (\bibinfo {year} {2019})},\ \Eprint
  {http://arxiv.org/abs/1812.08235} {arXiv:1812.08235 [hep-lat]} \BibitemShut
  {NoStop}%
\bibitem [{\citenamefont {Toublan}\ and\ \citenamefont
  {Kogut}(2005)}]{Toublan:2004ks}%
  \BibitemOpen
  \bibfield  {author} {\bibinfo {author} {\bibfnamefont {D.}~\bibnamefont
  {Toublan}}\ and\ \bibinfo {author} {\bibfnamefont {J.~B.}\ \bibnamefont
  {Kogut}},\ }\href {\doibase 10.1016/j.physletb.2004.11.018} {\bibfield
  {journal} {\bibinfo  {journal} {Phys. Lett.}\ }\textbf {\bibinfo {volume}
  {B605}},\ \bibinfo {pages} {129} (\bibinfo {year} {2005})},\ \Eprint
  {http://arxiv.org/abs/hep-ph/0409310} {arXiv:hep-ph/0409310 [hep-ph]}
  \BibitemShut {NoStop}%
\bibitem [{\citenamefont {Endrodi}\ \emph {et~al.}(2011)\citenamefont
  {Endrodi}, \citenamefont {Fodor}, \citenamefont {Katz},\ and\ \citenamefont
  {Szabo}}]{Endrodi:2011gv}%
  \BibitemOpen
  \bibfield  {author} {\bibinfo {author} {\bibfnamefont {G.}~\bibnamefont
  {Endrodi}}, \bibinfo {author} {\bibfnamefont {Z.}~\bibnamefont {Fodor}},
  \bibinfo {author} {\bibfnamefont {S.~D.}\ \bibnamefont {Katz}}, \ and\
  \bibinfo {author} {\bibfnamefont {K.~K.}\ \bibnamefont {Szabo}},\ }\href
  {\doibase 10.1007/JHEP04(2011)001} {\bibfield  {journal} {\bibinfo  {journal}
  {JHEP}\ }\textbf {\bibinfo {volume} {04}},\ \bibinfo {pages} {001} (\bibinfo
  {year} {2011})},\ \Eprint {http://arxiv.org/abs/1102.1356} {arXiv:1102.1356
  [hep-lat]} \BibitemShut {NoStop}%
\bibitem [{\citenamefont {Bellwied}\ \emph {et~al.}(2015)\citenamefont
  {Bellwied}, \citenamefont {Borsanyi}, \citenamefont {Fodor}, \citenamefont
  {Günther}, \citenamefont {Katz}, \citenamefont {Ratti},\ and\ \citenamefont
  {Szabo}}]{Bellwied:2015rza}%
  \BibitemOpen
  \bibfield  {author} {\bibinfo {author} {\bibfnamefont {R.}~\bibnamefont
  {Bellwied}}, \bibinfo {author} {\bibfnamefont {S.}~\bibnamefont {Borsanyi}},
  \bibinfo {author} {\bibfnamefont {Z.}~\bibnamefont {Fodor}}, \bibinfo
  {author} {\bibfnamefont {J.}~\bibnamefont {Günther}}, \bibinfo {author}
  {\bibfnamefont {S.~D.}\ \bibnamefont {Katz}}, \bibinfo {author}
  {\bibfnamefont {C.}~\bibnamefont {Ratti}}, \ and\ \bibinfo {author}
  {\bibfnamefont {K.~K.}\ \bibnamefont {Szabo}},\ }\href {\doibase
  10.1016/j.physletb.2015.11.011} {\bibfield  {journal} {\bibinfo  {journal}
  {Phys. Lett.}\ }\textbf {\bibinfo {volume} {B751}},\ \bibinfo {pages} {559}
  (\bibinfo {year} {2015})},\ \Eprint {http://arxiv.org/abs/1507.07510}
  {arXiv:1507.07510 [hep-lat]} \BibitemShut {NoStop}%
\bibitem [{\citenamefont {Giordano}\ \emph
  {et~al.}(2020{\natexlab{a}})\citenamefont {Giordano}, \citenamefont {Kapas},
  \citenamefont {Katz}, \citenamefont {Nogradi},\ and\ \citenamefont
  {Pasztor}}]{Giordano:2020uvk}%
  \BibitemOpen
  \bibfield  {author} {\bibinfo {author} {\bibfnamefont {M.}~\bibnamefont
  {Giordano}}, \bibinfo {author} {\bibfnamefont {K.}~\bibnamefont {Kapas}},
  \bibinfo {author} {\bibfnamefont {S.~D.}\ \bibnamefont {Katz}}, \bibinfo
  {author} {\bibfnamefont {D.}~\bibnamefont {Nogradi}}, \ and\ \bibinfo
  {author} {\bibfnamefont {A.}~\bibnamefont {Pasztor}},\ }\href {\doibase
  10.1103/PhysRevD.102.034503} {\bibfield  {journal} {\bibinfo  {journal}
  {Phys. Rev. D}\ }\textbf {\bibinfo {volume} {102}},\ \bibinfo {pages}
  {034503} (\bibinfo {year} {2020}{\natexlab{a}})},\ \Eprint
  {http://arxiv.org/abs/2003.04355} {arXiv:2003.04355 [hep-lat]} \BibitemShut
  {NoStop}%
\bibitem [{\citenamefont {Giordano}\ \emph
  {et~al.}(2020{\natexlab{b}})\citenamefont {Giordano}, \citenamefont {Kapas},
  \citenamefont {Katz}, \citenamefont {Nogradi},\ and\ \citenamefont
  {Pasztor}}]{Giordano:2020roi}%
  \BibitemOpen
  \bibfield  {author} {\bibinfo {author} {\bibfnamefont {M.}~\bibnamefont
  {Giordano}}, \bibinfo {author} {\bibfnamefont {K.}~\bibnamefont {Kapas}},
  \bibinfo {author} {\bibfnamefont {S.~D.}\ \bibnamefont {Katz}}, \bibinfo
  {author} {\bibfnamefont {D.}~\bibnamefont {Nogradi}}, \ and\ \bibinfo
  {author} {\bibfnamefont {A.}~\bibnamefont {Pasztor}},\ }\href {\doibase
  10.1007/JHEP05(2020)088} {\bibfield  {journal} {\bibinfo  {journal} {JHEP}\
  }\textbf {\bibinfo {volume} {05}},\ \bibinfo {pages} {088} (\bibinfo {year}
  {2020}{\natexlab{b}})},\ \Eprint {http://arxiv.org/abs/2004.10800}
  {arXiv:2004.10800 [hep-lat]} \BibitemShut {NoStop}%
\bibitem [{\citenamefont {Scherzer}\ \emph {et~al.}(2020)\citenamefont
  {Scherzer}, \citenamefont {Sexty},\ and\ \citenamefont
  {Stamatescu}}]{Scherzer:2020kiu}%
  \BibitemOpen
  \bibfield  {author} {\bibinfo {author} {\bibfnamefont {M.}~\bibnamefont
  {Scherzer}}, \bibinfo {author} {\bibfnamefont {D.}~\bibnamefont {Sexty}}, \
  and\ \bibinfo {author} {\bibfnamefont {I.-O.}\ \bibnamefont {Stamatescu}},\
  }\href {\doibase 10.1103/PhysRevD.102.014515} {\bibfield  {journal} {\bibinfo
   {journal} {Phys. Rev. D}\ }\textbf {\bibinfo {volume} {102}},\ \bibinfo
  {pages} {014515} (\bibinfo {year} {2020})},\ \Eprint
  {http://arxiv.org/abs/2004.05372} {arXiv:2004.05372 [hep-lat]} \BibitemShut
  {NoStop}%
\bibitem [{\citenamefont {Attanasio}\ \emph {et~al.}(2020)\citenamefont
  {Attanasio}, \citenamefont {J\"ager},\ and\ \citenamefont
  {Ziegler}}]{Attanasio:2020spv}%
  \BibitemOpen
  \bibfield  {author} {\bibinfo {author} {\bibfnamefont {F.}~\bibnamefont
  {Attanasio}}, \bibinfo {author} {\bibfnamefont {B.}~\bibnamefont {J\"ager}},
  \ and\ \bibinfo {author} {\bibfnamefont {F.~P.}\ \bibnamefont {Ziegler}},\
  }\href {\doibase 10.1140/epja/s10050-020-00256-z} {\bibfield  {journal}
  {\bibinfo  {journal} {Eur. Phys. J. A}\ }\textbf {\bibinfo {volume} {56}},\
  \bibinfo {pages} {251} (\bibinfo {year} {2020})},\ \Eprint
  {http://arxiv.org/abs/2006.00476} {arXiv:2006.00476 [hep-lat]} \BibitemShut
  {NoStop}%
\bibitem [{\citenamefont {P\'asztor}\ \emph {et~al.}(2021)\citenamefont
  {P\'asztor}, \citenamefont {Sz\'ep},\ and\ \citenamefont
  {Mark\'o}}]{Pasztor:2020dur}%
  \BibitemOpen
  \bibfield  {author} {\bibinfo {author} {\bibfnamefont {A.}~\bibnamefont
  {P\'asztor}}, \bibinfo {author} {\bibfnamefont {Z.}~\bibnamefont {Sz\'ep}}, \
  and\ \bibinfo {author} {\bibfnamefont {G.}~\bibnamefont {Mark\'o}},\ }\href
  {\doibase 10.1103/PhysRevD.103.034511} {\bibfield  {journal} {\bibinfo
  {journal} {Phys. Rev. D}\ }\textbf {\bibinfo {volume} {103}},\ \bibinfo
  {pages} {034511} (\bibinfo {year} {2021})},\ \Eprint
  {http://arxiv.org/abs/2010.00394} {arXiv:2010.00394 [hep-lat]} \BibitemShut
  {NoStop}%
\bibitem [{\citenamefont {Fischer}\ and\ \citenamefont
  {Luecker}(2013)}]{Fischer:2012vc}%
  \BibitemOpen
  \bibfield  {author} {\bibinfo {author} {\bibfnamefont {C.~S.}\ \bibnamefont
  {Fischer}}\ and\ \bibinfo {author} {\bibfnamefont {J.}~\bibnamefont
  {Luecker}},\ }\href {\doibase 10.1016/j.physletb.2012.11.054} {\bibfield
  {journal} {\bibinfo  {journal} {Phys. Lett.}\ }\textbf {\bibinfo {volume}
  {B718}},\ \bibinfo {pages} {1036} (\bibinfo {year} {2013})},\ \Eprint
  {http://arxiv.org/abs/1206.5191} {arXiv:1206.5191 [hep-ph]} \BibitemShut
  {NoStop}%
\bibitem [{\citenamefont {Fischer}\ \emph {et~al.}(2014)\citenamefont
  {Fischer}, \citenamefont {Luecker},\ and\ \citenamefont
  {Welzbacher}}]{Fischer:2014ata}%
  \BibitemOpen
  \bibfield  {author} {\bibinfo {author} {\bibfnamefont {C.~S.}\ \bibnamefont
  {Fischer}}, \bibinfo {author} {\bibfnamefont {J.}~\bibnamefont {Luecker}}, \
  and\ \bibinfo {author} {\bibfnamefont {C.~A.}\ \bibnamefont {Welzbacher}},\
  }\href {\doibase 10.1103/PhysRevD.90.034022} {\bibfield  {journal} {\bibinfo
  {journal} {Phys. Rev.}\ }\textbf {\bibinfo {volume} {D90}},\ \bibinfo {pages}
  {034022} (\bibinfo {year} {2014})},\ \Eprint {http://arxiv.org/abs/1405.4762}
  {arXiv:1405.4762 [hep-ph]} \BibitemShut {NoStop}%
\bibitem [{\citenamefont {Isserstedt}\ \emph {et~al.}(2019)\citenamefont
  {Isserstedt}, \citenamefont {Buballa}, \citenamefont {Fischer},\ and\
  \citenamefont {Gunkel}}]{Isserstedt:2019pgx}%
  \BibitemOpen
  \bibfield  {author} {\bibinfo {author} {\bibfnamefont {P.}~\bibnamefont
  {Isserstedt}}, \bibinfo {author} {\bibfnamefont {M.}~\bibnamefont {Buballa}},
  \bibinfo {author} {\bibfnamefont {C.~S.}\ \bibnamefont {Fischer}}, \ and\
  \bibinfo {author} {\bibfnamefont {P.~J.}\ \bibnamefont {Gunkel}},\ }\href
  {\doibase 10.1103/PhysRevD.100.074011} {\bibfield  {journal} {\bibinfo
  {journal} {Phys. Rev. D}\ }\textbf {\bibinfo {volume} {100}},\ \bibinfo
  {pages} {074011} (\bibinfo {year} {2019})},\ \Eprint
  {http://arxiv.org/abs/1906.11644} {arXiv:1906.11644 [hep-ph]} \BibitemShut
  {NoStop}%
\bibitem [{\citenamefont {Bratovic}\ \emph {et~al.}(2013)\citenamefont
  {Bratovic}, \citenamefont {Hatsuda},\ and\ \citenamefont
  {Weise}}]{Bratovic:2012qs}%
  \BibitemOpen
  \bibfield  {author} {\bibinfo {author} {\bibfnamefont {N.~M.}\ \bibnamefont
  {Bratovic}}, \bibinfo {author} {\bibfnamefont {T.}~\bibnamefont {Hatsuda}}, \
  and\ \bibinfo {author} {\bibfnamefont {W.}~\bibnamefont {Weise}},\ }\href
  {\doibase 10.1016/j.physletb.2013.01.003} {\bibfield  {journal} {\bibinfo
  {journal} {Phys. Lett.}\ }\textbf {\bibinfo {volume} {B719}},\ \bibinfo
  {pages} {131} (\bibinfo {year} {2013})},\ \Eprint
  {http://arxiv.org/abs/1204.3788} {arXiv:1204.3788 [hep-ph]} \BibitemShut
  {NoStop}%
\bibitem [{\citenamefont {Kov\'acs}\ \emph {et~al.}(2016)\citenamefont
  {Kov\'acs}, \citenamefont {Sz\'ep},\ and\ \citenamefont
  {Wolf}}]{Kovacs:2016juc}%
  \BibitemOpen
  \bibfield  {author} {\bibinfo {author} {\bibfnamefont {P.}~\bibnamefont
  {Kov\'acs}}, \bibinfo {author} {\bibfnamefont {Z.}~\bibnamefont {Sz\'ep}}, \
  and\ \bibinfo {author} {\bibfnamefont {G.}~\bibnamefont {Wolf}},\ }\href
  {\doibase 10.1103/PhysRevD.93.114014} {\bibfield  {journal} {\bibinfo
  {journal} {Phys. Rev. D}\ }\textbf {\bibinfo {volume} {93}},\ \bibinfo
  {pages} {114014} (\bibinfo {year} {2016})},\ \Eprint
  {http://arxiv.org/abs/1601.05291} {arXiv:1601.05291 [hep-ph]} \BibitemShut
  {NoStop}%
\bibitem [{\citenamefont {Fu}\ \emph {et~al.}(2020)\citenamefont {Fu},
  \citenamefont {Pawlowski},\ and\ \citenamefont {Rennecke}}]{Fu:2019hdw}%
  \BibitemOpen
  \bibfield  {author} {\bibinfo {author} {\bibfnamefont {W.-j.}\ \bibnamefont
  {Fu}}, \bibinfo {author} {\bibfnamefont {J.~M.}\ \bibnamefont {Pawlowski}}, \
  and\ \bibinfo {author} {\bibfnamefont {F.}~\bibnamefont {Rennecke}},\ }\href
  {\doibase 10.1103/PhysRevD.101.054032} {\bibfield  {journal} {\bibinfo
  {journal} {Phys. Rev.}\ }\textbf {\bibinfo {volume} {D101}},\ \bibinfo
  {pages} {054032} (\bibinfo {year} {2020})},\ \Eprint
  {http://arxiv.org/abs/1909.02991} {arXiv:1909.02991 [hep-ph]} \BibitemShut
  {NoStop}%
\bibitem [{\citenamefont {Gao}\ and\ \citenamefont
  {Pawlowski}(2020{\natexlab{a}})}]{Gao:2020fbl}%
  \BibitemOpen
  \bibfield  {author} {\bibinfo {author} {\bibfnamefont {F.}~\bibnamefont
  {Gao}}\ and\ \bibinfo {author} {\bibfnamefont {J.~M.}\ \bibnamefont
  {Pawlowski}},\ }\href@noop {} {\  (\bibinfo {year} {2020}{\natexlab{a}})},\
  \Eprint {http://arxiv.org/abs/2010.13705} {arXiv:2010.13705 [hep-ph]}
  \BibitemShut {NoStop}%
\bibitem [{\citenamefont {Gao}\ and\ \citenamefont
  {Pawlowski}(2020{\natexlab{b}})}]{Gao:2020qsj}%
  \BibitemOpen
  \bibfield  {author} {\bibinfo {author} {\bibfnamefont {F.}~\bibnamefont
  {Gao}}\ and\ \bibinfo {author} {\bibfnamefont {J.~M.}\ \bibnamefont
  {Pawlowski}},\ }\href {\doibase 10.1103/PhysRevD.102.034027} {\bibfield
  {journal} {\bibinfo  {journal} {Phys. Rev. D}\ }\textbf {\bibinfo {volume}
  {102}},\ \bibinfo {pages} {034027} (\bibinfo {year} {2020}{\natexlab{b}})},\
  \Eprint {http://arxiv.org/abs/2002.07500} {arXiv:2002.07500 [hep-ph]}
  \BibitemShut {NoStop}%
\bibitem [{\citenamefont {Andronic}\ \emph {et~al.}(2006)\citenamefont
  {Andronic}, \citenamefont {Braun-Munzinger},\ and\ \citenamefont
  {Stachel}}]{Andronic:2005yp}%
  \BibitemOpen
  \bibfield  {author} {\bibinfo {author} {\bibfnamefont {A.}~\bibnamefont
  {Andronic}}, \bibinfo {author} {\bibfnamefont {P.}~\bibnamefont
  {Braun-Munzinger}}, \ and\ \bibinfo {author} {\bibfnamefont {J.}~\bibnamefont
  {Stachel}},\ }\href {\doibase 10.1016/j.nuclphysa.2006.03.012} {\bibfield
  {journal} {\bibinfo  {journal} {Nucl. Phys. A}\ }\textbf {\bibinfo {volume}
  {772}},\ \bibinfo {pages} {167} (\bibinfo {year} {2006})},\ \Eprint
  {http://arxiv.org/abs/nucl-th/0511071} {arXiv:nucl-th/0511071} \BibitemShut
  {NoStop}%
\bibitem [{\citenamefont {Becattini}\ \emph {et~al.}(2013)\citenamefont
  {Becattini}, \citenamefont {Bleicher}, \citenamefont {Kollegger},
  \citenamefont {Schuster}, \citenamefont {Steinheimer},\ and\ \citenamefont
  {Stock}}]{Becattini:2012xb}%
  \BibitemOpen
  \bibfield  {author} {\bibinfo {author} {\bibfnamefont {F.}~\bibnamefont
  {Becattini}}, \bibinfo {author} {\bibfnamefont {M.}~\bibnamefont {Bleicher}},
  \bibinfo {author} {\bibfnamefont {T.}~\bibnamefont {Kollegger}}, \bibinfo
  {author} {\bibfnamefont {T.}~\bibnamefont {Schuster}}, \bibinfo {author}
  {\bibfnamefont {J.}~\bibnamefont {Steinheimer}}, \ and\ \bibinfo {author}
  {\bibfnamefont {R.}~\bibnamefont {Stock}},\ }\href {\doibase
  10.1103/PhysRevLett.111.082302} {\bibfield  {journal} {\bibinfo  {journal}
  {Phys. Rev. Lett.}\ }\textbf {\bibinfo {volume} {111}},\ \bibinfo {pages}
  {082302} (\bibinfo {year} {2013})},\ \Eprint {http://arxiv.org/abs/1212.2431}
  {arXiv:1212.2431 [nucl-th]} \BibitemShut {NoStop}%
\bibitem [{\citenamefont {Alba}\ \emph {et~al.}(2014)\citenamefont {Alba},
  \citenamefont {Alberico}, \citenamefont {Bellwied}, \citenamefont {Bluhm},
  \citenamefont {Mantovani~Sarti}, \citenamefont {Nahrgang},\ and\
  \citenamefont {Ratti}}]{Alba:2014eba}%
  \BibitemOpen
  \bibfield  {author} {\bibinfo {author} {\bibfnamefont {P.}~\bibnamefont
  {Alba}}, \bibinfo {author} {\bibfnamefont {W.}~\bibnamefont {Alberico}},
  \bibinfo {author} {\bibfnamefont {R.}~\bibnamefont {Bellwied}}, \bibinfo
  {author} {\bibfnamefont {M.}~\bibnamefont {Bluhm}}, \bibinfo {author}
  {\bibfnamefont {V.}~\bibnamefont {Mantovani~Sarti}}, \bibinfo {author}
  {\bibfnamefont {M.}~\bibnamefont {Nahrgang}}, \ and\ \bibinfo {author}
  {\bibfnamefont {C.}~\bibnamefont {Ratti}},\ }\href {\doibase
  10.1016/j.physletb.2014.09.052} {\bibfield  {journal} {\bibinfo  {journal}
  {Phys. Lett. B}\ }\textbf {\bibinfo {volume} {738}},\ \bibinfo {pages} {305}
  (\bibinfo {year} {2014})},\ \Eprint {http://arxiv.org/abs/1403.4903}
  {arXiv:1403.4903 [hep-ph]} \BibitemShut {NoStop}%
\bibitem [{\citenamefont {Vovchenko}\ \emph {et~al.}(2016)\citenamefont
  {Vovchenko}, \citenamefont {Begun},\ and\ \citenamefont
  {Gorenstein}}]{Vovchenko:2015idt}%
  \BibitemOpen
  \bibfield  {author} {\bibinfo {author} {\bibfnamefont {V.}~\bibnamefont
  {Vovchenko}}, \bibinfo {author} {\bibfnamefont {V.}~\bibnamefont {Begun}}, \
  and\ \bibinfo {author} {\bibfnamefont {M.}~\bibnamefont {Gorenstein}},\
  }\href {\doibase 10.1103/PhysRevC.93.064906} {\bibfield  {journal} {\bibinfo
  {journal} {Phys. Rev. C}\ }\textbf {\bibinfo {volume} {93}},\ \bibinfo
  {pages} {064906} (\bibinfo {year} {2016})},\ \Eprint
  {http://arxiv.org/abs/1512.08025} {arXiv:1512.08025 [nucl-th]} \BibitemShut
  {NoStop}%
\bibitem [{\citenamefont {Adamczyk}\ \emph {et~al.}(2017)\citenamefont
  {Adamczyk} \emph {et~al.}}]{Adamczyk:2017iwn}%
  \BibitemOpen
  \bibfield  {author} {\bibinfo {author} {\bibfnamefont {L.}~\bibnamefont
  {Adamczyk}} \emph {et~al.} (\bibinfo {collaboration} {STAR}),\ }\href
  {\doibase 10.1103/PhysRevC.96.044904} {\bibfield  {journal} {\bibinfo
  {journal} {Phys. Rev. C}\ }\textbf {\bibinfo {volume} {96}},\ \bibinfo
  {pages} {044904} (\bibinfo {year} {2017})},\ \Eprint
  {http://arxiv.org/abs/1701.07065} {arXiv:1701.07065 [nucl-ex]} \BibitemShut
  {NoStop}%
\bibitem [{\citenamefont {Bluhm}\ \emph {et~al.}(2020)\citenamefont {Bluhm},
  \citenamefont {Nahrgang},\ and\ \citenamefont {Pawlowski}}]{Bluhm:2020rha}%
  \BibitemOpen
  \bibfield  {author} {\bibinfo {author} {\bibfnamefont {M.}~\bibnamefont
  {Bluhm}}, \bibinfo {author} {\bibfnamefont {M.}~\bibnamefont {Nahrgang}}, \
  and\ \bibinfo {author} {\bibfnamefont {J.~M.}\ \bibnamefont {Pawlowski}},\
  }\href@noop {} {\  (\bibinfo {year} {2020})},\ \Eprint
  {http://arxiv.org/abs/2004.08608} {arXiv:2004.08608 [nucl-th]} \BibitemShut
  {NoStop}%
\bibitem [{\citenamefont {Aamodt}\ \emph
  {et~al.}(2010{\natexlab{a}})\citenamefont {Aamodt} \emph
  {et~al.}}]{Aamodt:2010pa}%
  \BibitemOpen
  \bibfield  {author} {\bibinfo {author} {\bibfnamefont {K.}~\bibnamefont
  {Aamodt}} \emph {et~al.} (\bibinfo {collaboration} {ALICE}),\ }\href
  {\doibase 10.1103/PhysRevLett.105.252302} {\bibfield  {journal} {\bibinfo
  {journal} {Phys. Rev. Lett.}\ }\textbf {\bibinfo {volume} {105}},\ \bibinfo
  {pages} {252302} (\bibinfo {year} {2010}{\natexlab{a}})},\ \Eprint
  {http://arxiv.org/abs/1011.3914} {arXiv:1011.3914 [nucl-ex]} \BibitemShut
  {NoStop}%
\bibitem [{\citenamefont {Aamodt}\ \emph
  {et~al.}(2011{\natexlab{a}})\citenamefont {Aamodt} \emph
  {et~al.}}]{Aamodt:2010jd}%
  \BibitemOpen
  \bibfield  {author} {\bibinfo {author} {\bibfnamefont {K.}~\bibnamefont
  {Aamodt}} \emph {et~al.} (\bibinfo {collaboration} {ALICE}),\ }\href
  {\doibase 10.1016/j.physletb.2010.12.020} {\bibfield  {journal} {\bibinfo
  {journal} {Phys. Lett. B}\ }\textbf {\bibinfo {volume} {696}},\ \bibinfo
  {pages} {30} (\bibinfo {year} {2011}{\natexlab{a}})},\ \Eprint
  {http://arxiv.org/abs/1012.1004} {arXiv:1012.1004 [nucl-ex]} \BibitemShut
  {NoStop}%
\bibitem [{\citenamefont {Chatrchyan}\ \emph {et~al.}(2011)\citenamefont
  {Chatrchyan} \emph {et~al.}}]{Chatrchyan:2011sx}%
  \BibitemOpen
  \bibfield  {author} {\bibinfo {author} {\bibfnamefont {S.}~\bibnamefont
  {Chatrchyan}} \emph {et~al.} (\bibinfo {collaboration} {CMS}),\ }\href
  {\doibase 10.1103/PhysRevC.84.024906} {\bibfield  {journal} {\bibinfo
  {journal} {Phys. Rev. C}\ }\textbf {\bibinfo {volume} {84}},\ \bibinfo
  {pages} {024906} (\bibinfo {year} {2011})},\ \Eprint
  {http://arxiv.org/abs/1102.1957} {arXiv:1102.1957 [nucl-ex]} \BibitemShut
  {NoStop}%
\bibitem [{\citenamefont {Aamodt}\ \emph
  {et~al.}(2011{\natexlab{b}})\citenamefont {Aamodt} \emph
  {et~al.}}]{Aamodt:2010cz}%
  \BibitemOpen
  \bibfield  {author} {\bibinfo {author} {\bibfnamefont {K.}~\bibnamefont
  {Aamodt}} \emph {et~al.} (\bibinfo {collaboration} {ALICE}),\ }\href
  {\doibase 10.1103/PhysRevLett.106.032301} {\bibfield  {journal} {\bibinfo
  {journal} {Phys. Rev. Lett.}\ }\textbf {\bibinfo {volume} {106}},\ \bibinfo
  {pages} {032301} (\bibinfo {year} {2011}{\natexlab{b}})},\ \Eprint
  {http://arxiv.org/abs/1012.1657} {arXiv:1012.1657 [nucl-ex]} \BibitemShut
  {NoStop}%
\bibitem [{\citenamefont {Aamodt}\ \emph
  {et~al.}(2010{\natexlab{b}})\citenamefont {Aamodt} \emph
  {et~al.}}]{Aamodt:2010pb}%
  \BibitemOpen
  \bibfield  {author} {\bibinfo {author} {\bibfnamefont {K.}~\bibnamefont
  {Aamodt}} \emph {et~al.} (\bibinfo {collaboration} {ALICE}),\ }\href
  {\doibase 10.1103/PhysRevLett.105.252301} {\bibfield  {journal} {\bibinfo
  {journal} {Phys. Rev. Lett.}\ }\textbf {\bibinfo {volume} {105}},\ \bibinfo
  {pages} {252301} (\bibinfo {year} {2010}{\natexlab{b}})},\ \Eprint
  {http://arxiv.org/abs/1011.3916} {arXiv:1011.3916 [nucl-ex]} \BibitemShut
  {NoStop}%
\bibitem [{\citenamefont {Aamodt}\ \emph
  {et~al.}(2011{\natexlab{c}})\citenamefont {Aamodt} \emph
  {et~al.}}]{ALICE:2011ab}%
  \BibitemOpen
  \bibfield  {author} {\bibinfo {author} {\bibfnamefont {K.}~\bibnamefont
  {Aamodt}} \emph {et~al.} (\bibinfo {collaboration} {ALICE}),\ }\href
  {\doibase 10.1103/PhysRevLett.107.032301} {\bibfield  {journal} {\bibinfo
  {journal} {Phys. Rev. Lett.}\ }\textbf {\bibinfo {volume} {107}},\ \bibinfo
  {pages} {032301} (\bibinfo {year} {2011}{\natexlab{c}})},\ \Eprint
  {http://arxiv.org/abs/1105.3865} {arXiv:1105.3865 [nucl-ex]} \BibitemShut
  {NoStop}%
\bibitem [{\citenamefont {Aamodt}\ \emph {et~al.}(2008)\citenamefont {Aamodt}
  \emph {et~al.}}]{Aamodt:2008zz}%
  \BibitemOpen
  \bibfield  {author} {\bibinfo {author} {\bibfnamefont {K.}~\bibnamefont
  {Aamodt}} \emph {et~al.} (\bibinfo {collaboration} {ALICE}),\ }\href
  {\doibase 10.1088/1748-0221/3/08/S08002} {\bibfield  {journal} {\bibinfo
  {journal} {JINST}\ }\textbf {\bibinfo {volume} {3}},\ \bibinfo {pages}
  {S08002} (\bibinfo {year} {2008})}\BibitemShut {NoStop}%
\bibitem [{\citenamefont {Mohanty}(2011)}]{Mohanty:2011nm}%
  \BibitemOpen
  \bibfield  {author} {\bibinfo {author} {\bibfnamefont {B.}~\bibnamefont
  {Mohanty}} (\bibinfo {collaboration} {STAR}),\ }\href {\doibase
  10.1088/0954-3899/38/12/124023} {\bibfield  {journal} {\bibinfo  {journal}
  {J. Phys. G}\ }\textbf {\bibinfo {volume} {38}},\ \bibinfo {pages} {124023}
  (\bibinfo {year} {2011})},\ \Eprint {http://arxiv.org/abs/1106.5902}
  {arXiv:1106.5902 [nucl-ex]} \BibitemShut {NoStop}%
\bibitem [{\citenamefont {Wilczek}(2011)}]{Wilczek:2010ae}%
  \BibitemOpen
  \bibfield  {author} {\bibinfo {author} {\bibfnamefont {F.}~\bibnamefont
  {Wilczek}},\ }\href {\doibase 10.1007/978-3-642-13293-3_1} {\bibfield
  {journal} {\bibinfo  {journal} {Lect. Notes Phys.}\ }\textbf {\bibinfo
  {volume} {814}},\ \bibinfo {pages} {1} (\bibinfo {year} {2011})},\ \Eprint
  {http://arxiv.org/abs/1001.2729} {arXiv:1001.2729 [hep-ph]} \BibitemShut
  {NoStop}%
\bibitem [{\citenamefont {Sorin}\ \emph {et~al.}(2011)\citenamefont {Sorin},
  \citenamefont {Kekelidze}, \citenamefont {Kovalenko}, \citenamefont
  {Lednicky}, \citenamefont {Meshkov},\ and\ \citenamefont
  {Trubnikov}}]{Sorin:2011zz}%
  \BibitemOpen
  \bibfield  {author} {\bibinfo {author} {\bibfnamefont {A.}~\bibnamefont
  {Sorin}}, \bibinfo {author} {\bibfnamefont {V.}~\bibnamefont {Kekelidze}},
  \bibinfo {author} {\bibfnamefont {A.}~\bibnamefont {Kovalenko}}, \bibinfo
  {author} {\bibfnamefont {R.}~\bibnamefont {Lednicky}}, \bibinfo {author}
  {\bibfnamefont {I.}~\bibnamefont {Meshkov}}, \ and\ \bibinfo {author}
  {\bibfnamefont {G.}~\bibnamefont {Trubnikov}},\ }\href {\doibase
  10.1016/j.nuclphysa.2011.02.118} {\bibfield  {journal} {\bibinfo  {journal}
  {Nucl. Phys. A}\ }\textbf {\bibinfo {volume} {855}},\ \bibinfo {pages} {510}
  (\bibinfo {year} {2011})}\BibitemShut {NoStop}%
\bibitem [{\citenamefont {Haque}\ \emph
  {et~al.}(2014{\natexlab{a}})\citenamefont {Haque}, \citenamefont
  {Bandyopadhyay}, \citenamefont {Andersen}, \citenamefont {Mustafa},
  \citenamefont {Strickland},\ and\ \citenamefont {Su}}]{Haque:2014rua}%
  \BibitemOpen
  \bibfield  {author} {\bibinfo {author} {\bibfnamefont {N.}~\bibnamefont
  {Haque}}, \bibinfo {author} {\bibfnamefont {A.}~\bibnamefont
  {Bandyopadhyay}}, \bibinfo {author} {\bibfnamefont {J.~O.}\ \bibnamefont
  {Andersen}}, \bibinfo {author} {\bibfnamefont {M.~G.}\ \bibnamefont
  {Mustafa}}, \bibinfo {author} {\bibfnamefont {M.}~\bibnamefont {Strickland}},
  \ and\ \bibinfo {author} {\bibfnamefont {N.}~\bibnamefont {Su}},\ }\href
  {\doibase 10.1007/JHEP05(2014)027} {\bibfield  {journal} {\bibinfo  {journal}
  {JHEP}\ }\textbf {\bibinfo {volume} {05}},\ \bibinfo {pages} {027} (\bibinfo
  {year} {2014}{\natexlab{a}})},\ \Eprint {http://arxiv.org/abs/1402.6907}
  {arXiv:1402.6907 [hep-ph]} \BibitemShut {NoStop}%
\bibitem [{\citenamefont {Andersen}\ \emph {et~al.}(1999)\citenamefont
  {Andersen}, \citenamefont {Braaten},\ and\ \citenamefont
  {Strickland}}]{Andersen:1999fw}%
  \BibitemOpen
  \bibfield  {author} {\bibinfo {author} {\bibfnamefont {J.~O.}\ \bibnamefont
  {Andersen}}, \bibinfo {author} {\bibfnamefont {E.}~\bibnamefont {Braaten}}, \
  and\ \bibinfo {author} {\bibfnamefont {M.}~\bibnamefont {Strickland}},\
  }\href {\doibase 10.1103/PhysRevLett.83.2139} {\bibfield  {journal} {\bibinfo
   {journal} {Phys. Rev. Lett.}\ }\textbf {\bibinfo {volume} {83}},\ \bibinfo
  {pages} {2139} (\bibinfo {year} {1999})},\ \Eprint
  {http://arxiv.org/abs/hep-ph/9902327} {arXiv:hep-ph/9902327} \BibitemShut
  {NoStop}%
\bibitem [{\citenamefont {Andersen}\ \emph
  {et~al.}(2000{\natexlab{a}})\citenamefont {Andersen}, \citenamefont
  {Braaten},\ and\ \citenamefont {Strickland}}]{Andersen:1999va}%
  \BibitemOpen
  \bibfield  {author} {\bibinfo {author} {\bibfnamefont {J.~O.}\ \bibnamefont
  {Andersen}}, \bibinfo {author} {\bibfnamefont {E.}~\bibnamefont {Braaten}}, \
  and\ \bibinfo {author} {\bibfnamefont {M.}~\bibnamefont {Strickland}},\
  }\href {\doibase 10.1103/PhysRevD.61.074016} {\bibfield  {journal} {\bibinfo
  {journal} {Phys. Rev. D}\ }\textbf {\bibinfo {volume} {61}},\ \bibinfo
  {pages} {074016} (\bibinfo {year} {2000}{\natexlab{a}})},\ \Eprint
  {http://arxiv.org/abs/hep-ph/9908323} {arXiv:hep-ph/9908323} \BibitemShut
  {NoStop}%
\bibitem [{\citenamefont {Andersen}\ \emph
  {et~al.}(2000{\natexlab{b}})\citenamefont {Andersen}, \citenamefont
  {Braaten},\ and\ \citenamefont {Strickland}}]{Andersen:2000zn}%
  \BibitemOpen
  \bibfield  {author} {\bibinfo {author} {\bibfnamefont {J.~O.}\ \bibnamefont
  {Andersen}}, \bibinfo {author} {\bibfnamefont {E.}~\bibnamefont {Braaten}}, \
  and\ \bibinfo {author} {\bibfnamefont {M.}~\bibnamefont {Strickland}},\
  }\href {\doibase 10.1103/PhysRevD.62.045004} {\bibfield  {journal} {\bibinfo
  {journal} {Phys. Rev. D}\ }\textbf {\bibinfo {volume} {62}},\ \bibinfo
  {pages} {045004} (\bibinfo {year} {2000}{\natexlab{b}})},\ \Eprint
  {http://arxiv.org/abs/hep-ph/0002048} {arXiv:hep-ph/0002048} \BibitemShut
  {NoStop}%
\bibitem [{\citenamefont {Andersen}\ \emph {et~al.}(2001)\citenamefont
  {Andersen}, \citenamefont {Braaten},\ and\ \citenamefont
  {Strickland}}]{Andersen:2000yj}%
  \BibitemOpen
  \bibfield  {author} {\bibinfo {author} {\bibfnamefont {J.~O.}\ \bibnamefont
  {Andersen}}, \bibinfo {author} {\bibfnamefont {E.}~\bibnamefont {Braaten}}, \
  and\ \bibinfo {author} {\bibfnamefont {M.}~\bibnamefont {Strickland}},\
  }\href {\doibase 10.1103/PhysRevD.63.105008} {\bibfield  {journal} {\bibinfo
  {journal} {Phys. Rev. D}\ }\textbf {\bibinfo {volume} {63}},\ \bibinfo
  {pages} {105008} (\bibinfo {year} {2001})},\ \Eprint
  {http://arxiv.org/abs/hep-ph/0007159} {arXiv:hep-ph/0007159} \BibitemShut
  {NoStop}%
\bibitem [{\citenamefont {Andersen}\ \emph {et~al.}(2002)\citenamefont
  {Andersen}, \citenamefont {Braaten}, \citenamefont {Petitgirard},\ and\
  \citenamefont {Strickland}}]{Andersen:2002ey}%
  \BibitemOpen
  \bibfield  {author} {\bibinfo {author} {\bibfnamefont {J.~O.}\ \bibnamefont
  {Andersen}}, \bibinfo {author} {\bibfnamefont {E.}~\bibnamefont {Braaten}},
  \bibinfo {author} {\bibfnamefont {E.}~\bibnamefont {Petitgirard}}, \ and\
  \bibinfo {author} {\bibfnamefont {M.}~\bibnamefont {Strickland}},\ }\href
  {\doibase 10.1103/PhysRevD.66.085016} {\bibfield  {journal} {\bibinfo
  {journal} {Phys. Rev. D}\ }\textbf {\bibinfo {volume} {66}},\ \bibinfo
  {pages} {085016} (\bibinfo {year} {2002})},\ \Eprint
  {http://arxiv.org/abs/hep-ph/0205085} {arXiv:hep-ph/0205085} \BibitemShut
  {NoStop}%
\bibitem [{\citenamefont {Andersen}\ and\ \citenamefont
  {Strickland}(2001)}]{Andersen:2001ez}%
  \BibitemOpen
  \bibfield  {author} {\bibinfo {author} {\bibfnamefont {J.~O.}\ \bibnamefont
  {Andersen}}\ and\ \bibinfo {author} {\bibfnamefont {M.}~\bibnamefont
  {Strickland}},\ }\href {\doibase 10.1103/PhysRevD.64.105012} {\bibfield
  {journal} {\bibinfo  {journal} {Phys. Rev. D}\ }\textbf {\bibinfo {volume}
  {64}},\ \bibinfo {pages} {105012} (\bibinfo {year} {2001})},\ \Eprint
  {http://arxiv.org/abs/hep-ph/0105214} {arXiv:hep-ph/0105214} \BibitemShut
  {NoStop}%
\bibitem [{\citenamefont {Andersen}\ \emph {et~al.}(2004)\citenamefont
  {Andersen}, \citenamefont {Petitgirard},\ and\ \citenamefont
  {Strickland}}]{Andersen:2003zk}%
  \BibitemOpen
  \bibfield  {author} {\bibinfo {author} {\bibfnamefont {J.~O.}\ \bibnamefont
  {Andersen}}, \bibinfo {author} {\bibfnamefont {E.}~\bibnamefont
  {Petitgirard}}, \ and\ \bibinfo {author} {\bibfnamefont {M.}~\bibnamefont
  {Strickland}},\ }\href {\doibase 10.1103/PhysRevD.70.045001} {\bibfield
  {journal} {\bibinfo  {journal} {Phys. Rev. D}\ }\textbf {\bibinfo {volume}
  {70}},\ \bibinfo {pages} {045001} (\bibinfo {year} {2004})},\ \Eprint
  {http://arxiv.org/abs/hep-ph/0302069} {arXiv:hep-ph/0302069} \BibitemShut
  {NoStop}%
\bibitem [{\citenamefont {Andersen}\ and\ \citenamefont
  {Strickland}(2005)}]{Andersen:2004fp}%
  \BibitemOpen
  \bibfield  {author} {\bibinfo {author} {\bibfnamefont {J.~O.}\ \bibnamefont
  {Andersen}}\ and\ \bibinfo {author} {\bibfnamefont {M.}~\bibnamefont
  {Strickland}},\ }\href {\doibase 10.1016/j.aop.2004.09.017} {\bibfield
  {journal} {\bibinfo  {journal} {Annals Phys.}\ }\textbf {\bibinfo {volume}
  {317}},\ \bibinfo {pages} {281} (\bibinfo {year} {2005})},\ \Eprint
  {http://arxiv.org/abs/hep-ph/0404164} {arXiv:hep-ph/0404164} \BibitemShut
  {NoStop}%
\bibitem [{\citenamefont {Andersen}\ \emph {et~al.}(2009)\citenamefont
  {Andersen}, \citenamefont {Strickland},\ and\ \citenamefont
  {Su}}]{Andersen:2009tw}%
  \BibitemOpen
  \bibfield  {author} {\bibinfo {author} {\bibfnamefont {J.~O.}\ \bibnamefont
  {Andersen}}, \bibinfo {author} {\bibfnamefont {M.}~\bibnamefont
  {Strickland}}, \ and\ \bibinfo {author} {\bibfnamefont {N.}~\bibnamefont
  {Su}},\ }\href {\doibase 10.1103/PhysRevD.80.085015} {\bibfield  {journal}
  {\bibinfo  {journal} {Phys. Rev. D}\ }\textbf {\bibinfo {volume} {80}},\
  \bibinfo {pages} {085015} (\bibinfo {year} {2009})},\ \Eprint
  {http://arxiv.org/abs/0906.2936} {arXiv:0906.2936 [hep-ph]} \BibitemShut
  {NoStop}%
\bibitem [{\citenamefont {Andersen}\ \emph
  {et~al.}(2010{\natexlab{a}})\citenamefont {Andersen}, \citenamefont
  {Strickland},\ and\ \citenamefont {Su}}]{Andersen:2009tc}%
  \BibitemOpen
  \bibfield  {author} {\bibinfo {author} {\bibfnamefont {J.~O.}\ \bibnamefont
  {Andersen}}, \bibinfo {author} {\bibfnamefont {M.}~\bibnamefont
  {Strickland}}, \ and\ \bibinfo {author} {\bibfnamefont {N.}~\bibnamefont
  {Su}},\ }\href {\doibase 10.1103/PhysRevLett.104.122003} {\bibfield
  {journal} {\bibinfo  {journal} {Phys. Rev. Lett.}\ }\textbf {\bibinfo
  {volume} {104}},\ \bibinfo {pages} {122003} (\bibinfo {year}
  {2010}{\natexlab{a}})},\ \Eprint {http://arxiv.org/abs/0911.0676}
  {arXiv:0911.0676 [hep-ph]} \BibitemShut {NoStop}%
\bibitem [{\citenamefont {Andersen}\ \emph
  {et~al.}(2010{\natexlab{b}})\citenamefont {Andersen}, \citenamefont
  {Strickland},\ and\ \citenamefont {Su}}]{Andersen:2010ct}%
  \BibitemOpen
  \bibfield  {author} {\bibinfo {author} {\bibfnamefont {J.~O.}\ \bibnamefont
  {Andersen}}, \bibinfo {author} {\bibfnamefont {M.}~\bibnamefont
  {Strickland}}, \ and\ \bibinfo {author} {\bibfnamefont {N.}~\bibnamefont
  {Su}},\ }\href {\doibase 10.1007/JHEP08(2010)113} {\bibfield  {journal}
  {\bibinfo  {journal} {JHEP}\ }\textbf {\bibinfo {volume} {08}},\ \bibinfo
  {pages} {113} (\bibinfo {year} {2010}{\natexlab{b}})},\ \Eprint
  {http://arxiv.org/abs/1005.1603} {arXiv:1005.1603 [hep-ph]} \BibitemShut
  {NoStop}%
\bibitem [{\citenamefont {Andersen}\ \emph
  {et~al.}(2011{\natexlab{a}})\citenamefont {Andersen}, \citenamefont
  {Leganger}, \citenamefont {Strickland},\ and\ \citenamefont
  {Su}}]{Andersen:2010wu}%
  \BibitemOpen
  \bibfield  {author} {\bibinfo {author} {\bibfnamefont {J.~O.}\ \bibnamefont
  {Andersen}}, \bibinfo {author} {\bibfnamefont {L.~E.}\ \bibnamefont
  {Leganger}}, \bibinfo {author} {\bibfnamefont {M.}~\bibnamefont
  {Strickland}}, \ and\ \bibinfo {author} {\bibfnamefont {N.}~\bibnamefont
  {Su}},\ }\href {\doibase 10.1016/j.physletb.2010.12.070} {\bibfield
  {journal} {\bibinfo  {journal} {Phys. Lett. B}\ }\textbf {\bibinfo {volume}
  {696}},\ \bibinfo {pages} {468} (\bibinfo {year} {2011}{\natexlab{a}})},\
  \Eprint {http://arxiv.org/abs/1009.4644} {arXiv:1009.4644 [hep-ph]}
  \BibitemShut {NoStop}%
\bibitem [{\citenamefont {Andersen}\ \emph
  {et~al.}(2011{\natexlab{b}})\citenamefont {Andersen}, \citenamefont
  {Leganger}, \citenamefont {Strickland},\ and\ \citenamefont
  {Su}}]{Andersen:2011sf}%
  \BibitemOpen
  \bibfield  {author} {\bibinfo {author} {\bibfnamefont {J.~O.}\ \bibnamefont
  {Andersen}}, \bibinfo {author} {\bibfnamefont {L.~E.}\ \bibnamefont
  {Leganger}}, \bibinfo {author} {\bibfnamefont {M.}~\bibnamefont
  {Strickland}}, \ and\ \bibinfo {author} {\bibfnamefont {N.}~\bibnamefont
  {Su}},\ }\href {\doibase 10.1007/JHEP08(2011)053} {\bibfield  {journal}
  {\bibinfo  {journal} {JHEP}\ }\textbf {\bibinfo {volume} {08}},\ \bibinfo
  {pages} {053} (\bibinfo {year} {2011}{\natexlab{b}})},\ \Eprint
  {http://arxiv.org/abs/1103.2528} {arXiv:1103.2528 [hep-ph]} \BibitemShut
  {NoStop}%
\bibitem [{\citenamefont {Andersen}\ \emph
  {et~al.}(2011{\natexlab{c}})\citenamefont {Andersen}, \citenamefont
  {Leganger}, \citenamefont {Strickland},\ and\ \citenamefont
  {Su}}]{Andersen:2011ug}%
  \BibitemOpen
  \bibfield  {author} {\bibinfo {author} {\bibfnamefont {J.~O.}\ \bibnamefont
  {Andersen}}, \bibinfo {author} {\bibfnamefont {L.~E.}\ \bibnamefont
  {Leganger}}, \bibinfo {author} {\bibfnamefont {M.}~\bibnamefont
  {Strickland}}, \ and\ \bibinfo {author} {\bibfnamefont {N.}~\bibnamefont
  {Su}},\ }\href {\doibase 10.1103/PhysRevD.84.087703} {\bibfield  {journal}
  {\bibinfo  {journal} {Phys. Rev. D}\ }\textbf {\bibinfo {volume} {84}},\
  \bibinfo {pages} {087703} (\bibinfo {year} {2011}{\natexlab{c}})},\ \Eprint
  {http://arxiv.org/abs/1106.0514} {arXiv:1106.0514 [hep-ph]} \BibitemShut
  {NoStop}%
\bibitem [{\citenamefont {Haque}\ \emph
  {et~al.}(2013{\natexlab{a}})\citenamefont {Haque}, \citenamefont {Mustafa},\
  and\ \citenamefont {Strickland}}]{Haque:2012my}%
  \BibitemOpen
  \bibfield  {author} {\bibinfo {author} {\bibfnamefont {N.}~\bibnamefont
  {Haque}}, \bibinfo {author} {\bibfnamefont {M.~G.}\ \bibnamefont {Mustafa}},
  \ and\ \bibinfo {author} {\bibfnamefont {M.}~\bibnamefont {Strickland}},\
  }\href {\doibase 10.1103/PhysRevD.87.105007} {\bibfield  {journal} {\bibinfo
  {journal} {Phys. Rev. D}\ }\textbf {\bibinfo {volume} {87}},\ \bibinfo
  {pages} {105007} (\bibinfo {year} {2013}{\natexlab{a}})},\ \Eprint
  {http://arxiv.org/abs/1212.1797} {arXiv:1212.1797 [hep-ph]} \BibitemShut
  {NoStop}%
\bibitem [{\citenamefont {Mogliacci}\ \emph {et~al.}(2013)\citenamefont
  {Mogliacci}, \citenamefont {Andersen}, \citenamefont {Strickland},
  \citenamefont {Su},\ and\ \citenamefont {Vuorinen}}]{Mogliacci:2013mca}%
  \BibitemOpen
  \bibfield  {author} {\bibinfo {author} {\bibfnamefont {S.}~\bibnamefont
  {Mogliacci}}, \bibinfo {author} {\bibfnamefont {J.~O.}\ \bibnamefont
  {Andersen}}, \bibinfo {author} {\bibfnamefont {M.}~\bibnamefont
  {Strickland}}, \bibinfo {author} {\bibfnamefont {N.}~\bibnamefont {Su}}, \
  and\ \bibinfo {author} {\bibfnamefont {A.}~\bibnamefont {Vuorinen}},\ }\href
  {\doibase 10.1007/JHEP12(2013)055} {\bibfield  {journal} {\bibinfo  {journal}
  {JHEP}\ }\textbf {\bibinfo {volume} {12}},\ \bibinfo {pages} {055} (\bibinfo
  {year} {2013})},\ \Eprint {http://arxiv.org/abs/1307.8098} {arXiv:1307.8098
  [hep-ph]} \BibitemShut {NoStop}%
\bibitem [{\citenamefont {Haque}\ \emph
  {et~al.}(2013{\natexlab{b}})\citenamefont {Haque}, \citenamefont {Mustafa},\
  and\ \citenamefont {Strickland}}]{Haque:2013qta}%
  \BibitemOpen
  \bibfield  {author} {\bibinfo {author} {\bibfnamefont {N.}~\bibnamefont
  {Haque}}, \bibinfo {author} {\bibfnamefont {M.~G.}\ \bibnamefont {Mustafa}},
  \ and\ \bibinfo {author} {\bibfnamefont {M.}~\bibnamefont {Strickland}},\
  }\href {\doibase 10.1007/JHEP07(2013)184} {\bibfield  {journal} {\bibinfo
  {journal} {JHEP}\ }\textbf {\bibinfo {volume} {07}},\ \bibinfo {pages} {184}
  (\bibinfo {year} {2013}{\natexlab{b}})},\ \Eprint
  {http://arxiv.org/abs/1302.3228} {arXiv:1302.3228 [hep-ph]} \BibitemShut
  {NoStop}%
\bibitem [{\citenamefont {Haque}\ \emph
  {et~al.}(2014{\natexlab{b}})\citenamefont {Haque}, \citenamefont {Andersen},
  \citenamefont {Mustafa}, \citenamefont {Strickland},\ and\ \citenamefont
  {Su}}]{Haque:2013sja}%
  \BibitemOpen
  \bibfield  {author} {\bibinfo {author} {\bibfnamefont {N.}~\bibnamefont
  {Haque}}, \bibinfo {author} {\bibfnamefont {J.~O.}\ \bibnamefont {Andersen}},
  \bibinfo {author} {\bibfnamefont {M.~G.}\ \bibnamefont {Mustafa}}, \bibinfo
  {author} {\bibfnamefont {M.}~\bibnamefont {Strickland}}, \ and\ \bibinfo
  {author} {\bibfnamefont {N.}~\bibnamefont {Su}},\ }\href {\doibase
  10.1103/PhysRevD.89.061701} {\bibfield  {journal} {\bibinfo  {journal} {Phys.
  Rev. D}\ }\textbf {\bibinfo {volume} {89}},\ \bibinfo {pages} {061701}
  (\bibinfo {year} {2014}{\natexlab{b}})},\ \Eprint
  {http://arxiv.org/abs/1309.3968} {arXiv:1309.3968 [hep-ph]} \BibitemShut
  {NoStop}%
\bibitem [{\citenamefont {Andersen}\ \emph {et~al.}(2016)\citenamefont
  {Andersen}, \citenamefont {Haque}, \citenamefont {Mustafa},\ and\
  \citenamefont {Strickland}}]{Andersen:2015eoa}%
  \BibitemOpen
  \bibfield  {author} {\bibinfo {author} {\bibfnamefont {J.~O.}\ \bibnamefont
  {Andersen}}, \bibinfo {author} {\bibfnamefont {N.}~\bibnamefont {Haque}},
  \bibinfo {author} {\bibfnamefont {M.~G.}\ \bibnamefont {Mustafa}}, \ and\
  \bibinfo {author} {\bibfnamefont {M.}~\bibnamefont {Strickland}},\ }\href
  {\doibase 10.1103/PhysRevD.93.054045} {\bibfield  {journal} {\bibinfo
  {journal} {Phys. Rev. D}\ }\textbf {\bibinfo {volume} {93}},\ \bibinfo
  {pages} {054045} (\bibinfo {year} {2016})},\ \Eprint
  {http://arxiv.org/abs/1511.04660} {arXiv:1511.04660 [hep-ph]} \BibitemShut
  {NoStop}%
\bibitem [{\citenamefont {Du}\ \emph {et~al.}(2020)\citenamefont {Du},
  \citenamefont {Strickland}, \citenamefont {Tantary},\ and\ \citenamefont
  {Zhang}}]{Du:2020odw}%
  \BibitemOpen
  \bibfield  {author} {\bibinfo {author} {\bibfnamefont {Q.}~\bibnamefont
  {Du}}, \bibinfo {author} {\bibfnamefont {M.}~\bibnamefont {Strickland}},
  \bibinfo {author} {\bibfnamefont {U.}~\bibnamefont {Tantary}}, \ and\
  \bibinfo {author} {\bibfnamefont {B.-W.}\ \bibnamefont {Zhang}},\ }\href
  {\doibase 10.1007/JHEP09(2020)038} {\bibfield  {journal} {\bibinfo  {journal}
  {JHEP}\ }\textbf {\bibinfo {volume} {09}},\ \bibinfo {pages} {038} (\bibinfo
  {year} {2020})},\ \Eprint {http://arxiv.org/abs/2006.02617} {arXiv:2006.02617
  [hep-ph]} \BibitemShut {NoStop}%
\bibitem [{\citenamefont {Braaten}\ and\ \citenamefont
  {Pisarski}(1992)}]{Braaten:1991gm}%
  \BibitemOpen
  \bibfield  {author} {\bibinfo {author} {\bibfnamefont {E.}~\bibnamefont
  {Braaten}}\ and\ \bibinfo {author} {\bibfnamefont {R.~D.}\ \bibnamefont
  {Pisarski}},\ }\href {\doibase 10.1103/PhysRevD.45.R1827} {\bibfield
  {journal} {\bibinfo  {journal} {Phys. Rev.}\ }\textbf {\bibinfo {volume}
  {D45}},\ \bibinfo {pages} {R1827} (\bibinfo {year} {1992})}\BibitemShut
  {NoStop}%
\end{thebibliography}%

\end{document}